\documentclass[twocolumn,twoside,slac_two]{revtex4}
\usepackage{graphicx}
\usepackage{fancyhdr}
\begin{document}

\title{Twelve Years of Education and Public Outreach with the Fermi Gamma-ray Space Telescope}

%

\author{L. Cominsky, K. McLin, A. Simonnet and the  Fermi E/PO team}
\affiliation{Sonoma State University, Rohnert Park, CA 94928, USA}

\begin{abstract}
During the past twelve years, NASA's {\it Fermi Gamma-ray Space Telescope} has supported a wide range of Education and Public Outreach (E/PO) activities, targeting K-14 students and the general public. The purpose of the {\it Fermi} E/PO program is to increase student and public understanding of the science of the high-energy Universe, through inspiring, engaging and educational activities linked to the mission's science objectives. The E/PO program has additional more general goals, including increasing the diversity of students in the Science, Technology, Engineering and Mathematics (STEM) pipeline, and increasing public awareness and understanding of {\it Fermi} science and technology.  {\it Fermi}'s multi-faceted E/PO program includes elements in each major outcome category:
Higher Education; Elementary and Secondary Education; Informal Education and Public Outreach.
\end{abstract}

\maketitle

\thispagestyle{fancy}


The Sonoma State University (SSU) Education and Public Outreach (E/PO) group leads the {\it Fermi} Education and Public Outreach program. Since 1999, we have produced multi-media products that have reached tens of millions, trained tens of thousands of teachers, distributed many thousands of educational handouts and done authentic research with hundreds of students. In the sections below, we will review some of the accomplishments during the past twelve years and will also present some future plans for the program.

\section{Higher Education}
{\it Fermi}  E/PO promotes STEM careers through the use of NASA data including research experiences for students and teachers (Global Telescope Network), education through STEM curriculum development projects (Big Ideas in Cosmology curriculum) and through enrichment activities ({\it e.g.,} Large Area Telescope simulator).

\subsection{Global Telescope Network}
The Global Telescope Network (GTN) provides students with the opportunity to engage in authentic research experiences; similar efforts have been shown to help students develop the type of scientific inquiry skills that are needed for future STEM-based careers. Over 200 high school and college students have used the GTN since 2004.  The GTN now consists of 38 partners from small college, high school and amateur observatories from around the world. {\it Fermi} E/PO staff have developed a NASA-approved website ({\tt http://gtn.sonoma.edu}) and regularly assist students to observe active galaxies and other targets using our robotic 14-inch telescope, GORT. GORT (the {\it GLAST} Optical Robotic Telescope) is located about 45 minutes northeast of the SSU campus, in a higher, darker and less foggy location. The GTN also provides access to southern hemisphere telescopes through the University of North Carolina's PROMPT cluster in Chile. WestEd interviews of four female high school GTN interns noted that ``students said that their interest in astronomy, science, and the scientific method had grown  considerably over the summer and (all) expressed interest in careers in science." The {\it Fermi} E/PO program is  continuing to support the GTN during FY13-16 in the areas of pipeline maintenance, work with minority institutions and with high school students.

\subsection{Big Ideas in Cosmology Curriculum}
In these challenging times, many universities are turning to online curricula, which are thought to improve learning as well as efficiency. Along with additional funding through the NASA ROSES program, {\it Fermi} E/PO is supporting the development of an online curriculum for college students: {\it The Big Ideas in Cosmology}. The potential audience for this course is conservatively estimated at 25,000 students per year. The evaluation plan includes extensive formative pedagogical evaluation by WestEd, expert scientific reviewers, and a diverse group of pilot testers. Publisher Kendall Hunt is supporting market testing and online implementation. We are currently field testing the first of three five-chapter modules ({\it Our Place in the Universe: Space and Time}) that comprise the course, while the other two modules ({\it The Darker Side of the Universe: Gravity, Black Holes and Dark Matter } and {\it Our Evolving Universe: Past, Present and Future}) are being developed in parallel.

\subsection{Large Area Telescope Simulator}
The Large Area Telescope (LAT) Simulator is aimed at students in grades 11-14, as well as at the general public. This interactive computer-based educational activity shows the results when gamma-ray photons of different energies and incident angles hit the LAT, Fermi's main scientific instrument. Originally sited at the SLAC Virtual Visitor's Center, the simulator is now at:
{\tt http://fermi.sonoma.edu/multimedia/latsim/}.

\section{Elementary and Secondary Education}
The {\it Fermi} E/PO program promotes inquiry into topics that are included in the National Science Education Content Standards A, B, \& D for grades 7-14, including forces and motion, the structure and evolution of the Universe, and the relationship between energy and matter. Many {\it Fermi} E/PO products also align to National Mathematics standards. All curricular materials have been approved by NASA Product Review, and are available free of charge. {\it Fermi} E/PO links the science objectives of the mission to well-tested, customer-focused and standards-aligned classroom materials (Black Hole Resources, Active Galaxy Education Unit and Pop-up book, TOPS guides, Supernova Education Unit). These materials have been distributed through (Educator Ambassador and on-line) teacher training workshops and through programs involving under-represented students (after-school clubs and Astro 4 Girls).

\subsection{Black Hole Resources}
SSU has created a fact sheet explaining black holes in both English and Spanish. The English version has been distributed to nearly 25,000 viewers, while the Spanish version has been distributed to more than 3400. The brochures were also distributed to attendees at the NASA/NSF Black Holes museum exhibit designed by the education group at Harvard. The complete black hole resource area can be found at: {\tt http://fermi.sonoma.edu/teachers/blackholes/}. The site includes links to multi-media shows, an educator's guide, teacher's workshop materials, printed materials, and links to external sites.

\subsection{Active Galaxy Education Unit}
\begin{figure*}[t]
\centering
\includegraphics[width=135mm]{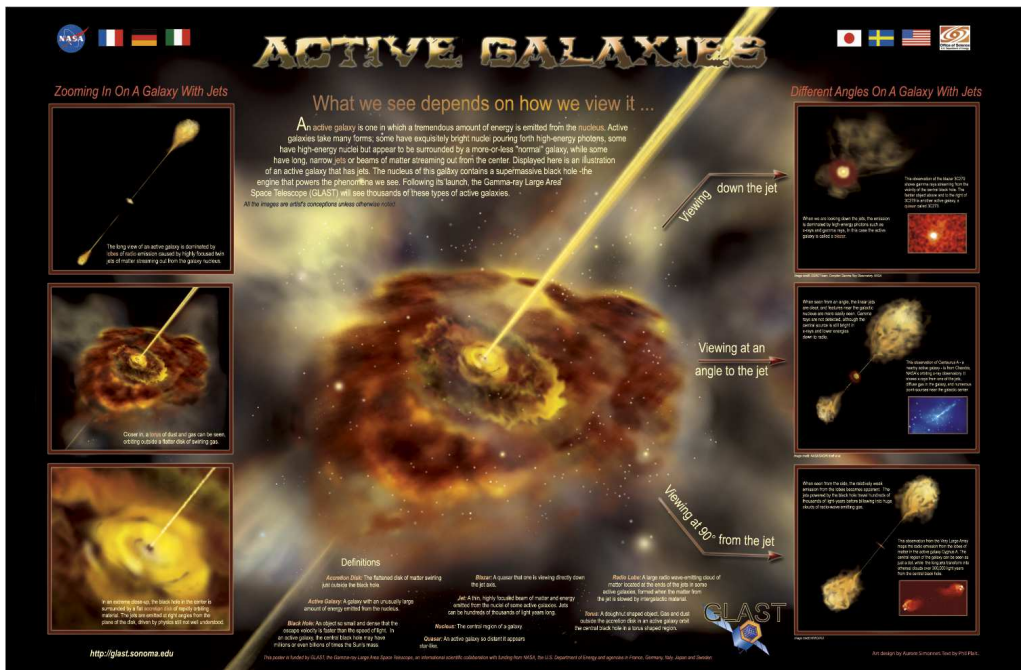}
\caption{Educational Wallsheet from the Active Galaxy Education Unit} \label{agn}
\end{figure*}

The original Active Galaxy Education Unit was developed for students in grades 9-14, and includes an educator's guide as well as an educational wallsheet (see Figure~\ref{agn}) which is used in one of the activities. The activities use Active Galaxies to teach standards-based science and math concepts including geometrical perspective, the small angle approximation and light travel time. More than 2600 of these guides have been distributed through workshops. The guide, the artwork, and other associated resources can be downloaded from: {\tt http://fermi.sonoma.edu/teachers/agn.php}.

For younger students (grades 3-8), we have developed an Active Galaxy pop-up book that shows the nucleus of an active galaxy in three dimensions. It also includes a ``Just-So" story entitled ``How the Galaxy Got its Jets," the Tasty Active Galaxy Activity (modeled on the paper activity from the original education guide), and a glossary. The pop-up book is accompanied by an educator's guide that includes standards alignment information, lesson plans and additional background information about active galaxies and the pop-up components. Almost 5000 have been distributed through workshops and classroom visits. For more information: {\tt http://fermi.sonoma.edu/teachers/popup.php}.

\subsection{TOPS guides}
There are three TOPS guides: {\it Far Out Math, Scale the Universe, {\rm and} Pi in the Sky}. All three guides use {\it Fermi} (then called {\it GLAST}) science to illustrate mathematical concepts for grades 5-12. The guides were developed through a sub-award to Ron Marson, TOPS Learning Systems, with technical and scientific oversight by {\it Fermi} E/PO. To date, we have distributed over 3000 copies of {\it Far Out Math}, almost 4000 copies of {\it Scale the Universe}, and over 2000 copies of {\it Pi in the Sky} through workshops for educators.

Using {\it Far Out Math}, students measure, scale, graph and problem solve, using examples derived from {\it Fermi}. They compare quantities as orders of magnitude, become familiar with scientific notation, and develop a concrete understanding of exponents and logarithms.

The size and scale of the Universe are among the most difficult concepts for students to understand. The {\it Scale the Universe} booklet  gently takes students through the orders-of-magnitude of the Universe starting with the familiar human scale, then going down to the infinitesimal and out to the astronomical.

The activities in {\it Pi in the Sky} teach about radians and degrees, angles, parallax, and apparent sizes of objects. Students measure the apparent sizes of objects at various distances and learn how we can measure the sizes of distant objects.  All three booklets can be downloaded from: {\tt http://fermi.sonoma.edu/teachers/tops.php}

\subsection{Supernova Education Unit}
The Supernova Education Unit is a set of four activities for grades 7-14 which uses supernovae and pulsars to teach standards-based science concepts including the electromagnetic spectrum, motion, magnetic fields, and science literacy. It was developed jointly with the {\it XMM-Newton} E/PO program. The Supernova Education Unit features four curriculum enhancement activities, background information, assessment information, extension and transfer activities, and detailed information about the physical science and mathematics content standards. One of the activities is available both as a paper/pencil and electronic image/spreadsheet activity. We have printed and distributed nearly 1000 guides and/or CDs through educator workshops. (The CDs include useful software, files that can be used to do the exercises and a PDF file of the educator guide.) The unit can be downloaded from {\tt http://xmm.sonoma.edu/edu/supernova/}.

\subsection{Educator Ambassador Program}
Educator Ambassadors (EAs) are master teachers who engage other educators in professional development that is directly connected to teaching practices. The EA program currently has 16 educators who annually train thousands of other educators at local, state, regional, and national teachers' meetings, giving SSU-approved workshops across the country. Every other year, they attend a week-long workshop at SSU: the most recent training was held in July 2012, and included a two-day mini-course about Gravitation. Previous mini-courses focused on the dark universe and particle physics. To date, the {\it Fermi}-funded EAs, together with SSU E/PO personnel, have directly trained over 31,000 students and teachers through over 370 training events. The EA program, as a whole, has directly trained over 57,000 teachers and students across North America since 2001.  The program is demonstrably successful, as evidenced by an evaluation of data from more than 1500 responses (140 workshops) compiled by WestEd. WestEd has also evaluated all five bi-annual EA trainings, and many workshops conducted by EAs and by SSU personnel. {\it Fermi} E/PO is supporting 10 EAs during FY13-16 including: Jeffrey Adkins (California), Teena Della (Canada), Michiel Ford (Kansas), Mandy Frantti (Michigan), Mary Garrett (Georgia), Bruce Hemp (Virginia), Christine Royce (Pennsylvania), Linda Smith (New Jersey), Daryl Taylor (Connecticut), and Pamela Whiffen (Arizona). The next training will be held in July 2014. For more information about the EA program, to download training materials and videos, or to meet the EAs, see {\tt http://epo.sonoma.edu/ea}.

\subsection{NASA's Multiwavelength Universe}
In 2011, the pilot course {\it NASA's Multiwavelength Universe} was developed, hosted on SSU servers, and Prof. Cominsky served as the Instructor of Record for a diverse group of 25 teachers. The course met the needs of the (predominantly middle-school) teachers by helping them gain content and pedagogical knowledge, as well as experiences with inquiry-based activities and use of the Internet to find NASA resources. The program was evaluated via a pre- and post-test, by collecting feedback from the participants, and also by WestEd. {\it Fermi} E/PO continued to support this course during 2012, and plans to continue to support the course through 2016. When the course is not in session, the course materials can be viewed and downloaded at: {\tt http://universe.sonoma.edu/cosmo/course}

 {\tt /view.php?id=5}.

\subsection{After-school clubs and Astro 4 Girls}
From 2004-2011, {\it Fermi, Swift, {\rm and} XMM-Newton} E/PO programs sponsored four different after-school clubs (grades 1-3, 4-6, 6-8 and 9-12) at three different Sonoma county public, predominantly Hispanic schools. The total number of students served each week during 2011 through these clubs was about 100.

In 2011, the Astrophysics Forum began a collaborative effort to reach middle-school girls through workshops to be held at libraries in March of each year. In 2012, two of the nine venues were served by {\it Fermi}-supported Educator Ambassadors Teena Della and Pam Whiffen. The need for this type of program is well documented, as girls continue to be under-represented in STEM fields, and in NASA technical careers. In addition, libraries have a need for authoritative and up-to-date technical information. {\it Fermi} E/PO plans to continue to support these workshops during 2013-2016.

\section{Informal Education and Public Outreach}
{\it Fermi} E/PO engages the public in sharing the experience of exploration and discovery through high-leverage multi-media experiences (Black Holes planetarium and PBS NOVA shows), through popular websites (Gamma-ray Burst Skymap, Epo's Chronicles), social media ({\it e.g.,} Facebook), interactive web-based activities (Space Mysteries, Einstein@Home), activities by amateur astronomers nation-wide (Supernova! toolkit) and inspirational printed materials.

\subsection{Black Holes: The Other Side of Infinity and Monster of the Milky Way}
In January 2006 the planetarium show {\it Black Holes: The Other Side of Infinity}, premiered at the Denver Museum of Nature \& Science. Jointly funded by NSF and NASA's {\it GLAST} (now {\it Fermi}) E/PO program, this full-dome digital show continues to be popular. The narrator is actor Liam Neeson, and the show was directed by Tom Lucas. To date, it has been shown in over 125 planetaria world-wide to over two million viewers in 16 different languages. Elements of the planetarium show were used as part of a PBS NOVA episode, {\it Monster of the Milky Way} (also directed by Tom Lucas). Both shows take a sweeping look at the many lives of black holes,  from studies of the supermassive black hole in the center of our Milky Way Galaxy to the birth of black holes in gamma-ray bursts and their seeming ubiquity in the universe. Prof. Cominsky was a science advisor for both shows. Initially viewed by over 10 million people (and rerun many times since its premiere in 2006) and accompanied by additional educational resources, it is available online at: {\tt http://www.pbs.org/wgbh/nova/space/}

{\tt monster-milky-way.html}.

\subsection{E/PO websites}
{\it Fermi} E/PO websites include the Gamma-ray Burst (GRB) Skymap site and Epo's Chronicles, a weekly webcomic.  The GRB Skymap site is aimed at the scientifically attentive public, including many amateur astronomers who ``chase" GRBs and is now supported by both {\it Swift} and {\it Fermi}. The GRB Skymap site includes short writeups of every GRB seen by {\it Fermi}, {\it Swift} and other satellites, as well as positional information and finding charts. The purpose is to have a complete record of every GRB observed since 2004, with user-friendly descriptions accessible to the scientifically attentive general public.  To visit the site: {\tt http://grb.sonoma.edu}.

Each week, SSU staff create, write, and draw the Epo's Chronicles web comic. It is then translated into French, Spanish and Italian. The comic is accompanied by additional background information, definitions of scientific terms, and links to multimedia resources. Monthly readership has doubled in the past year and the strip is now read by over 60,000 unique visitors each year.  Join Alkina and her sentient spaceship Epo, as they travel the galaxy trying to discover their origins and learning about space science: {\tt http://eposchronicles.org}.

\subsection{Social Media and Citizen Science}
{\it Fermi} social media pages on Facebook are regularly updated by SSU staff with the latest science news. We also support the {\it Fermi} iPhone app developed by mission partners and the {\it Fermi} Twitter feed @NASAFermi. During FY12 we began to partner with the NSF LIGO-funded citizen science project {\it Einstein@Home} to use {\it Fermi} data to search for gamma-ray pulsars.

\subsection{Interactive web-based Activities}
{\it Fermi} E/PO developed two  {\it Space Mysteries} - interactive games that teach space science: {\it Solar Supernova?} and {\it Galactic Doom?} Your mission in {\it Solar Supernova?} is to go through scientific notes taken by Professor Starzapoppin, examine the data he's collected from various stars, and see if Sun is going to eventually explode. In {\it Galactic Doom?}, Alkina (from Epo's Chronicles) narrates as students learn about the different shapes that galaxies take, what it means for a galaxy to be active, and how astronomers tell if galaxies are active by examining their images and spectra. Then students move on to observing some distant galaxies and making some of their own classifications. The final goal is to determine if our own galaxy is active, thus placing the Earth in danger. Play both {\it Space Mysteries}: {\tt http://mystery.sonoma.edu}.

Other interactive web-based activities include tours of the one-year {\it Fermi} LAT skymap using Google Sky and Microsoft's World-wide Telescope, and the {\it Fermi} Pulsar Explorer. The Pulsar Explorer is a Flash-based interactive map that includes information about more than 100 gamma-ray pulsars observed and/or discovered by {\it Fermi} since its launch in 2008. Categories include historical pulsars, {\it Fermi}-discovered gamma-ray pulsars, {\it Fermi}-detected radio pulsars, and {\it Fermi}-assisted radio pulsars. Clicking on symbols in the map brings up additional information about each pulsar. Originally developed to accompany a NASA media telecon about the pulsar discoveries, it can be viewed at: {\tt http://www.nasa.gov/externalflash/fermipulsar/}.

\subsection{Activities with Amateur Astronomers}
Together with the {\it Suzaku, Swift {\rm and} XMM-Newton} E/PO programs, and the Astronomical Society of the Pacific (ASP), {\it Fermi} E/PO helped develop  the Supernova! toolkit for the Night Sky Network (NSN). The toolkit includes instructions for activities for star parties, skymaps, videos, background information and PowerPoint slides. The NSN includes over 200 amateur astronomy clubs. Through 2011, the Supernova! toolkit has reached over 138,000 attendees through 1,284 events.  Of these events, 679 events reported including almost 25,000 minorities and over 39,000 women/girls. The individual activities in the toolkit are available for download through the NSN: {\tt http://nightsky.jpl.nasa.gov/}, click on ``Astronomy Activities" and then search for Supernova!

\subsection{Printed Materials}
{\it Fermi} has printed and distributed thousands of NASA-approved outreach materials including: the {\it Fermi} paper model, Race Card game, lithograph, and fact sheet.  The paper model includes complete instructions, definitions of {\it Fermi} components, and printed sheets needed to construct the model itself. The {\it Fermi} Race Card game is a strategy-oriented board game in which players (or teams of players) compete to build {\it Fermi} and then observe five different classes of objects. The {\it Fermi} litho is a one-page litho that describes the mission science, as well as providing instructions for the ``Build your own pulsar model" activity for students. The {\it Fermi} Fact Sheet is a four-page color brochure that describes the mission science in more detail than the litho, and has been updated to provide information on science highlights through the four years since launch. All these materials can be downloaded from: {\tt http://fermi.sonoma.edu/teachers/}.

A recent printed product is the two-year skymap poster, that calls out eight major discoveries published through 2010, and indicates their locations superimposed on a detailed map of the gamma-ray sky at energies greater than 100 MeV.   It can be downloaded from: {\tt http://fermi.sonoma.edu/multimedia/gallery/}. Many additional scientific illustrations can also be found at this same web location.

\bigskip

\section{Evaluation}
Since 2001, {\it Fermi} E/PO has supported a comprehensive program of external evaluation overseen by Dr. Edward Britton of the Math and Science program at WestEd. WestEd also is the external evaluator for SSU on the {\it Swift, NuSTAR} and {\it XMM-Newton} E/PO programs, as well as for the {\it Big Ideas in Cosmology} curriculum. WestEd has evaluated every educational resource produced by the E/PO team, as well as all five bi-annual Educator Ambassador trainings, many individual sessions conducted by the EAs, and the on-line NASA's Multi-wavelength Universe course. WestEd provides both formative and summative assessment that is critical to ensuring the high-quality of the {\it Fermi} E/PO program.

\bigskip 
\begin{acknowledgments}
The authors wish to thank the other members of the E/PO group at Sonoma State University, including Kevin John, Logan
Hill, Laura Chase and David McCall for their support. We also acknowledge contributions from former members of
the group including Tim Graves, Sarah Silva, Phil Plait, and Kamal Prasad.

This work has been supported by NASA NAS5-00147 through Stanford University.
\end{acknowledgments}

\end{document}